\documentclass[useAMS,usenatbib]{mn2e}
\usepackage{graphics}

\title[Effective temperatures of magnetic CP stars]
{Effective temperatures of magnetic CP stars from full spectral energy
  distributions} 

\author[\L. Lipski and K. St\c epie\'n]
 {\L. Lipski$^{1}$ and K. St\c epie\'n$^{1}$\\
$^{1}$Warsaw University Observatory, Al. Ujazdowskie 4, 00-478 Warszawa,
 Poland}

\date{Accepted --.
         Received -- ;
         in original form --}

\pubyear{2007}
\begin{document}
\maketitle
\label{firstpage}

\begin{abstract}

New determinations of effective temperatures of 23 magnetic, chemically
peculiar (mCP) stars were obtained from a fit of metal enhanced model
atmospheres to the observed spectral energy distributions (SED) from UV to
red. Temperatures of four more CP stars without magnetic measurements were
also obtained, of which three show light variations characteristic of
magnetic variables, and the fourth may possibly be normal. The
root-mean-square (RMS) method was used to fit the theoretical SED to the
observations corrected for reddening if necessary, with metallicity and
effective temperature as the fitting parameters. Gravity was assumed to be
equal to $\log g$ = 4 for main sequence stars and to $\log g$ = 3 for two
giants in the considered sample. Equal weights were given to the UV part
and visual part of SED. An attempt was made to obtain a three-parameter fit
(with reddening included) to the observed SED of HD 215441. It turned out
that the three-parameter fit gave spurious results due to correlations
among the parameters. We could determine the reddening of that star only by
{\em assuming} that the metallicity of the model is known, hence reducing the
number of fitting parameters to two.

Independently of the formal quality of fit resulting from the RMS method
applied to the whole SED, the quality of fit was additionally checked for each
star by determination of the temperature
from the best fitting model atmosphere to the UV part and the visual part
of SED separately. Both temperatures should be close to one another if the
global best fitting model satisfactorily describes the full observed
SED. This is the case for about a half of the investigated stars but the
difference exceeds 750 K for the remaining stars with the extreme values
above 2000 K. Possible reasons for such discrepancies are discussed.

New, revised calibrations of effective temperature and bolometric
corrections of mCP stars in terms of reddening free Str\" omgren indices
are given.

\end{abstract}

\section{Introduction}

Effective temperature, $T_e$, is the
stellar parameter of crucial astrophysical importance. It
needs to be accurately known prior to any detailed analysis of stellar
structure and evolution. According to definition, effective temperature is
determined from the integrated emergent energy flux which, in turn, can be
calculated from the observed energy distribution and angular diameter of
the star \citep{Code76}. The method cannot be used on a massive scale
because accurate observations of the energy distribution and angular
diameters are available for a limited number of stars only. Such stars
serve usually as standards for calibration of other temperature measures of
which most useful are color indices observed with different photometric
systems. To avoid problems with measuring angular diameters indirect
methods have been suggested, like the infrared flux method (IRFM) in which
the diameter can be calculated from the ratio of the observed to
(stellar) surface monochromatic
flux in infrared, assuming its weak dependence on
temperature \citep{BS77}. Since a large grid of realistic
model atmospheres computed by \citet{Kur79, Kur92} has became available (see
http://kurucz.harvard.edu./grids.html) effective temperature of a star can
be determined from a fit of a theoretical emergent flux to the observed
energy distribution. Model atmospheres are available for a broad range of
temperatures, gravities and different metal abundances (scaled solar
abundances). With effective temperature known, the angular diameter can be
calculated from the total observed energy flux. Because the energy
distribution depends only weakly on surface gravity, the simultaneous
determination of temperature and gravity is impractical, even in case of
stars with known metallicity.  Surface gravity is usually determined from
other data, e. g. from Balmer lines or spectral type.  
On the other hand, energy
distributions of two models with the same $T_e$ and $\log g$ but
different metal abundances may differ significantly. Metals produce a large
number of spectral lines in the UV, which block effectively a noticeable fraction of
the flux (this is called blanketing effect). The blocked flux is radiated
in longer wavelengths due to backwarming effect (an increase of
temperature of the corresponding atmospheric layers). Increased metal
abundance modifies the temperature distribution as a function of optical
depth in a sense that temperature is higher, compared to normal metal
abundance star, in layers where the visual part of the spectrum is formed
and it is lower in layers where the UV part of the spectrum is formed.  As
a result, calibrations based on visual energy distributions of normal stars
are not applicable to metal rich stars as they give systematically too high
effective temperatures. Stars with chemical composition deviating
significantly from solar should be calibrated separately.

Several groups of chemically peculiar (CP) stars with non-solar abundances exist
among upper main sequence stars \citep{Pre74}. One of them is a group of
magnetic CP (mCP) stars. They show a presence of strong,
ordered surface magnetic fields with intensities up to several kiloGauss
\citep{BBM03}, associated with large atmospheric overabundances of heavy
elements, among them silicon, rare earth elements and iron peak elements
\citep{Pre74}. Compared to other peculiar stars, e. g. HgMn, $\lambda$ Boo
type or Am stars, mCP stars show the largest deviations of spectral energy
distribution (SED) from normal stars. The overabundant elements block
effectively the UV flux which is re-radiated in longer wavelengths,
similarly as in metal rich stars. However, a precise shape of SED of each
mCP star depends sensitively on its detailed chemical composition. In many such stars
elements are not evenly distributed over the atmosphere. Instead, they show
concentrations in some parts of the stellar surface (chemical spots) and
are nonuniformly distributed in the vertical direction \citep{Rya03}. All
this complicates enormously modeling of their atmospheres. Even if one
considers only a surface averaged model of mCP star, the correct approach
requires an iterative process in which the atmospheric parameters and
chemical abundances are corrected at each step. Such a procedure has not
been so far applied to mCP stars. Several years ago there have been
attempts to model the observed energy distributions of a few stars with
individually tailored, but fixed, chemical compositions \citep{MS80, SM87}
but, due to a shortage of the modeling procedure, significant discrepancies
between the observed and computed SED remained. Very few individual SED of
chemically peculiar stars have been modeled since then (see
http://wwwuser.oat.ts.astro.it/castelli).

Many attempts to determine effective temperatures of mCP stars can be found
in literature. Unfortunately, most of them are based on photometry or, at
best, on visual parts of SED \citep{HN93, Sok98, AR00}. As a result they
are prone to systematic errors arising from peculiar SED and uncertain
reddening corrections. \citet{SD89} analyzed full SEDs (visual + UV from
IUE) but, in lack of metal enhanced models at that time, they fitted the
Kurucz solar composition models to the visual parts of SED and then they
calculated corrections resulting from the UV flux deficit to obtain final
effective temperatures.

On the other hand, a
large volume of photometric data in different photometric systems exist for
peculiar stars, which could be used to determine their temperatures if
reliable calibrations exist for them \citep{MD85, HN93, Nap93, HK96}.
A critical analysis of some of them is given by \citet{Ste94}.

The purpose of the present paper is to determine effective temperatures of
mCP stars for which UV and visual scans exist, by fitting metal enhanced
model atmospheres to dereddened (if necessary) SED. Thus determined
temperatures can be used to improve existing calibrations of photometric
color indices.  Selection criteria for including a star into our analysis
and stellar observational data are
described in the next section together with the method of obtaining the
final observed SED. The fitting method of model atmospheres to the
observations and the results are given in Section 3 together with the
discussion and new calibrations in terms of the color indices from
Str\" omgren photometry. The last section summarizes the main conclusions.

\begin{table*}
\begin{minipage}{175mm}
\caption[]{List of investigated stars.}
\label{list}
\begin{tabular}{rlcccccrrccccc} 
\hline
%\footnote{one additional scan of HD 40312 from Glushnieva et al. (1998a) 
%and two more scans from  Alekseeva et al. (1997) were also included}
HD & Other name & Pec. & $V$& $E(B-V)$& [$u-b$]& [$c_1$] & $d$(pc)& $n$(UV)&  $n$(Ad) &  
$n$(Br) &  $n$(Bu) & $n$(Gl) &    $n$(Kh) \\
\hline
15089 & $\iota$ Cas & Sr &4.53&0& 1.350 & 0.866&  43& 12 & 9 & &1&&1 \\
19832 & 56 Ari & Si & 5.77&0& 0.786 & 0.560 &114&90 & 23 & 2 &&& 1          \\
23387 & HII 717 & CrSi & 7.19&0& 1.269 & 0.917 &184& 2 &&&&& 1           \\
25823 & 41 Tau & SrSi & 5.17&0& 0.717 & 0.501 &152&4  &  15 & 2 & 1 & 1 &   \\
26571 & V1137 Tau & Si & 6.15&0.27& 0.720 & 0.512 &316&11 &&&&& 1          \\
27309 & 56 Tau & SiCr & 5.35&0& 0.892 & 0.559& 97&4 & 4 & 1 &&&         \\
34452 & IQ Aur & Si & 5.37&0& 0.648 & 0.374 &137&10 & 24 & 1 &&& 1 \\
37470 & BD-06 1274 & Si & 8.23&0.15& 0.892 & 0.668 &265&3 && 1 &&&   \\
40312 & $\theta$ Aur\footnote{Three more visual scans from \citet{Ale96, 
Ale97}  and \citet{Glu98a} were also included.}
 & Si & 2.62&0& 1.239 & 0.974 &53&23 & 20 &1 & 2 & 1 & 1 \\
43819 & V1155 Ori & Si & 6.27&0& 1.019 & 0.762 &193&1 & 5 & 1 &&&\\
65339 & 53 Cam & SrEuCr & 6.02&0& 1.299 & 0.739 &98&24 & 14 &&&&\\
92664 & V364 Car & Si & 5.50&0& 0.583 & 0.403 &143&10 && 1 &&&\\
98664 & $\sigma$ Leo & Si & 4.04&0& 1.259 & 1.018 &66&5 &&& 2 & 1 & 1\\
107966 & 13 Com & Am? & 5.18&0& 1.487 & 1.103 &87&6 &&&1 & 1 &\\
108662 & 17 Com A & SrCrEu & 5.24&0& 1.289 & 0.891 &83&7 & 4 & 1 & 1 & 1 &\\
108945 & 21 Com & Sr & 5.44&0& 1.509 & 1.087 &95&34 & 4 & 1 &&&\\
112413 & $\alpha^2$ CVn & EuSiCr & 2.90&0& 0.976 & 0.641 &34&34 &29&2 & 1 & 1 & 1\\
118022 & 78 Vir & Cr EuSr & 4.91&0& 1.395 & 0.938 &56&22 & 15 & 2 & 2 & 1 & 1\\
120198 & 84 UMa & EuCr & 5.68&0& 1.260 & 0.898 &86&4 & 5 && && 1 \\
124224 & CU Vir & Si & 5.01&0& 0.832 & 0.599 &80&4 & 51 & 1 & 1 & 1 & \\
125248 & CS Vir & EuCr & 5.86&0& 1.364 & 0.926 &90&7 & 15 & 0 &&&\\
133029 & BX Boo & SiCrSr & 6.35&0& 1.147 & 0.794 &146&4 & 15 & 1 &&&\\
144844 & HR 6003 & Si He-w & 5.86&0.12& 0.844 & 0.587 &131&2 && 1 &&&\\
152107 & 52 Her & SrCrEu &4.82&0& 1.383 & 0.933 &54& 8 && 2 & 1 & 1 &\\
171782 & QV Ser & SiCrEu & 7.85&0.17& 0.999 & 0.742 &286&2 & 4 &&&&\\
196502 & 73 Dra & SrCrEu & 5.19&0& 1.530 & 1.061 &128&2 & 24 &&&&\\
215441 & GL Lac & Si & 8.81&0.26& 0.568 & 0.315 &714&16 & 15 &&&&\\
\hline
\multicolumn{14}{l}{Ad - \citet{Ade89}, Br - \citet{Bre76}, Bu -
\citet{Bur85}, Gl - \citet{Glu98b}, Kh - \citet{Kha88}}
\end{tabular}
\end{minipage}
\end{table*}

\section{Selection of stars and data sources}

A star had to fulfill simultaneously the following criteria to be included
into further analysis:

\begin{itemize}
\item to be classified as chemically peculiar of CP2-type \citep{Pre74},

\item to be observed at least once by IUE with the large aperture, in low
dispersion and with both, short-wavelength (SW) and long-wavelength (LW) 
cameras,

\item to be observed spectrophotometrically in visual.
\end{itemize}

As a first step we folded the IUE archive with 5225 objects for which
required observations exist with the {\it General Catalog of Ap and Am
stars} by \citet{Ren91}. This step was necessary
because not all CP stars observed with IUE are marked as such in the
archive. 301 stars appear in both catalogs. They were checked against the
following catalogs of visual scans: \citet{Ade89},\citet{Ale96, Ale97},
\citet{Bre76}, \citet{Bur85}, \citet{Glu98a, Glu98b} and \citet{Kha88}. 
Observations exist for 137 stars from the
previous sample in at least one of the visual catalogs. These stars were
checked individually star by star and all non-CP2 stars were rejected. The
final sample contains 27 stars. Apart from well known, classical mCP stars
a few additional objects were included due to their peculiar character
mentioned in literature. A closer analysis at a later stage of our
investigation showed that not all of them belong to mCP stars but,
nevertheless, we obtained their effective temperatures.

Table~\ref{list} lists all the investigated stars with their apparent
peculiarities, magnitudes, accepted reddening, Str\" omgren reddening free
indices and distances. Last six columns give the numbers of UV scans from
IUE and visual scans from the respective catalogs used in the further
analysis. Note that HD 192678, considered by \citet{SD89}, is missing
here. This is because contradictory information on IUE observations of this
star exists. In the {\it Final merged log of IUE observations} ({\it CDS
Catalog VI/110}) HD 192678 appears with spectra SWP 7568 and LWR 6547
(which were used by \citealt{SD89}). However, according to the {\it INES
Catalog} (see below), these spectra are attached to HD 192679 which is only
about 0.3 of a magnitude brighter than HD 192678 and has a spectral type of
F5V. Apparent similarity of both stars might have caused the confusion. The
Simbad database gives no mention of IUE observations for either star. We
decided to exclude HD 192678 from our sample.

\subsection{IUE Observations}

The UV observations were extracted from {\it IUE Newly Extracted Spectra}
(INES) {\it Catalog} in which the flux in absolute units (cgs) is given
every 1.67 A for SW cameras and every 2.67 A for LW cameras. Observations
were reduced with the use of the New Spectral Image Processing System
(NEWSISP) method \citep{Gar97}. It is assumed, when reducing the
observations, that those obtained with the large aperture contain a full
stellar flux whereas only an unknown part of the flux is registered through
the small aperture because of the vignetting effect. The precise
calibration factor converting the small amplitude flux to the large
aperture flux must be determined individually for each scan. Without large
aperture observations such an individual factor cannot be found, so
stars with only small amplitude spectra were excluded from our analysis.

As a first step, we checked all large aperture spectra for internal
consistency. The following scans turned out to be discrepant: one SW scan
of HD 19832 (out of 39 listed in archive), two SW scans of HD 34452 (out of
7), and 2 LW scans of HD 98664 (out of 3). In the last case the LW scans
were additionally checked against SW and visual scans.  All discrepant
scans were rejected. Next, the average SED was formed from individual scans
weighted with inverse square root of the observational error which is
listed in the archive together with each measurement. Measurements flagged
with negative ``errors'' are outliers and were not included into the
average. Scaling factors for small aperture spectra were then determined
from a comparison of a median value of each small aperture spectrum to a
median value of the average large aperture spectrum. After multiplying all
small aperture spectra by the corresponding scaling factors, they were
averaged in the same way as those obtained with the large amplitude and
compared with them. In no case any systematic difference between small and
large amplitude spectra was detected, hence both sets were used to obtain
the (weighted) grand average SED. Only one SW and one LW spectrum is
available for HD 43819 and, unfortunately, the LW spectrum badly disagrees
with the SW and visual scans. The discrepant spectrum was rejected but the
star was retained. This is the only star in our analysis for which the full
(i. e. UV plus visual) SED is not available. In the last step we calculated
the values of the observed flux at the wavelengths for which model values
are listed by Kurucz. The latter values are given every 10 A for
wavelengths shorter than 2900 A and every 20 A for the longer ones. After
dividing the observed spectral range into intervals with centers
corresponding to the wavelengths in which model values of fluxes are given,
the weighted average value for each interval was calculated. Fluxes in the
overlapping wavelength interval were calculated from observations taken
with both cameras, allowing for the respective errors. Thus
obtained SEDs can directly be compared with theoretical models.

\subsection{Visual observations}

\citet{Ade89} list several scans for each observed star. The
spectral passband varies between 25 and 30 A and the number of measurements
per scan varies between 30 and 50. Fluxes are given in magnitudes and
always centered at nearly the same wavelengths, although small differences
among the scans are present. As a first step in forming the final SED
we checked the scans of each star for internal consistency. For two stars:
HD 171782 and HD 215441, additional scans corrected for interstellar
absorption are also listed. We excluded them from further
analysis because we applied the corresponding corrections to the whole SED
of these stars (see below). Otherwise, the scans did not show
discrepancies. Next, we reduced flux values of each star to a uniform set of
wavelengths using linear interpolation. For example, if most of the
observations were obtained for the wavelength 5556 but in some scans values
for 5550 A are given, they were interpolated to the former
wavelength. Then, mean SEDs were calculated assuming equal weights of all
measurements. Scans of variable stars were obtained in different phases of
the variation period so the mean SED can be treated as period
averaged, particularly when a substantial number of scans is available. As
a next step, the observations were converted into absolute units, using
Vega calibration \citep{BG04}

\begin{equation}
F_{\lambda} = F_{\rm{Vega,5556}}\left(\frac{5556^2}{\lambda^2}\right)
10^{-0.4(V-V_{\rm{Vega}} + m_{\lambda} - m_{5556})}\,,
\end{equation}

where $\lambda$ is the wavelength in A, $F_{\lambda}$ is the stellar flux
in absolute units (ergcm$^{-2}$s$^{-1}$A$^{-1}$), $F_{\rm{Vega,5556}} =
3.46\times 10^{-9}$ ergcm$^{-2}$s$^{-1}$A$^{-1}$, $V$ and $V_{\rm{Vega}}$ = 0.026
are the magnitudes of the star and Vega in $V$-band of the $UBV$ system, and
$m_{\lambda}$ and $m_{5556}$ are the magnitudes of the star at the
wavelength $\lambda$ and at 5556 A.

If, instead of averaging magnitudes, we had transformed them into
fluxes, the resulting
differences in the mean values turned out to be negligible (less than 0.5
\%), compared to the procedure described above.

\citet{Bre76} lists observations obtained by many different observers and
with different instrumentation. Bandwidths used by them extend from 10 to
100 A. For eleven stars the catalog contains only one scan and
for 5 of them two scans per star, extending over different spectral regions,
e.g. 3000 - 5500 A and 5000 - 8000 A. The analyzed scans
were first compared to those from the Adelman et al. catalog (if the
latter were available). Only in case of one scan of HD 112413 a substantial
discrepancy occurred. This scan was also discordant
with the UV observations so we rejected it. All the accepted scans were
then reduced to absolute units in the same way as those from Adelman et al.
catalog and merged into one SED by averaging measurements from
overlapping spectral intervals if two scans were available.

In the catalogs of \citet{Ale97}, \citet{Bur85}, 
\citet{Glu98a, Glu98b} and \citet{Kha88} (henceforth
ABGK catalogs) from 1 to 7 scans per star are given. \citet{Ale97} 
used the bandwidth of 100 A. We could not find any information
on bandwidths used by the other authors. Nearly all the  observations extend
from 3200 to about 8000 A but \citet{Ale97} and \citet{Glu98a} 
obtained observations up to 10800 A. For consistency with other
observations we ignored observations beyond 8500 A. The fluxes listed in
the ABGK catalogs are given in absolute units per different
length units. To reduce them to the common scale we applied
corresponding scaling factors. Next, fluxes of each star were
interpolated to a uniform wavelength set and averaged, assuming equal
weights.

In all the considered catalogs the observations obtained within the
Balmer lines or very close to them are present. Because we are interested
in continuum distribution only, the observations affected by Balmer lines
were rejected. The remaining continuum observations were interpolated to
the wavelengths listed in model computations.

Because the scans from \citet{Ade89} and \citet{Bre76} are given on the
same scale, a direct comparison of them for each star can be made.  No
systematic difference was found between both sets of scans for stars listed
in the catalogs. A good agreement means that the temperatures obtained for
stars not observed by \citet{Ade89} but present in the Breger catalog do
not need any systematic corrections. Scans from the ABGK catalogs could not
be directly compared with those from Adelman et al. or Breger catalogs, so
they were analyzed separately to keep track of any possible systematic
differences.  The results showed that they also agree well with the Adelman
et al. and Breger catalogs (see below).

\subsection{IUE + visual SED}

When a UV spectrum is plotted together with a visual scan, they usually
do not merge smoothly. Instead, a gap is often visible between them in the
wavelength region around 3200 A covered by both sets of observations. Visual
observations were retained here because formal errors of the UV
observations in the overlapping region are substantially larger than those
of visual observations. In majority of cases the UV observations lie
slightly lower than visual but there are also cases when they agree well
and even some cases when they lie higher. Small shifts between UV and
visual observations result probably from calibration errors. No attempt was
made to remove them. 

The observed SED should be corrected for interstellar absorption before
fitting the theoretical SED. Unfortunately, standard methods of
determination the individual values of the interstellar reddening cannot be
applied to mCP stars due to their peculiar photometric indices
\citep{Wolff67, SM80}.  Other, indirect methods, e. g. based on maps of the
distribution of the interstellar matter, can also result in gross errors
because interstellar dust has a very patchy structure. Recent investigation
of interstellar absorption for many W UMa-type stars by \citet{RD97} showed
that the reddening of nearby stars had been significantly overestimated by
earlier authors. In fact, interstellar reddening is close to zero for stars
closer than 100 pc and rarely exceeds 0.03 for distances up to 200 pc (see
also \citealt{Fru94}). Based on these results we assumed zero reddening for
all stars closer than 100 pc. Each star lying further than 100 pc was
checked individually. If literature data or maps of interstellar matter
suggested $E(B-V) \le 0.03$ we assumed zero reddening.  We believe that the
overestimate of the reddening is as dangerous for the determination of true
SED as its underestimate, so in cases when no lower limit for the reddening
is given, it is safer to assume zero.  Several stars are, however,
significantly reddened and values of $E(B-V)$ up to 0.27 mag can be found
in literature. With one exception we accepted the literature values as they
are, although some of them are very uncertain. For one star, HD 215441, the
fit of the model atmosphere was so poor with the published $E(B-V) = 0.20$
that we decided to do a numerical experiment for this star by keeping the
reddening as an additional free parameter to be determined from the best
fit (see 3.1). The observed SEDs of reddened stars
were corrected for interstellar extinction using the curve given by
\citet{Car89}, see also \citet{FM07}. The accepted values of $E(B-V)$ are
listed in Table~\ref{list} with references given in the comments to
individual stars. As a final step before fitting theoretical SED, the
observed fluxes were converted to units used by Kurucz.

\begin{table}
\caption[]{Effective temperatures of stars common in Ad+Br and ABGK
catalogs.}
\label{common}
\begin{tabular}{rrrr} 
\hline
HD & $T_e$(Ad+Br) & $T_e$(ABGK) & $\Delta T_e$ \\
\hline
15089 & 8250 & 8500 & -250 \\
19832 & 12750 & 13000 & -250         \\
25823 & 13250 & 13500 & -250   \\
34452 & 14750 & 15000 & -250  \\
40312 & 10500 & 10250 & 250 \\
108662 & 10250 & 9750 & 500 \\
112413 & 11750 & 12000 & -250 \\
118022 & 9000 & 9250 & -250 \\
120198 & 10500 & 10500 & 0 \\
124224 & 13000 & 13250 & -250  \\
152107 & 8750 & 8500 & 250 \\
\hline
\end{tabular}
\end{table}

\begin{figure*}
\begin{center}
\rotatebox{0}{\scalebox{0.85}{\includegraphics{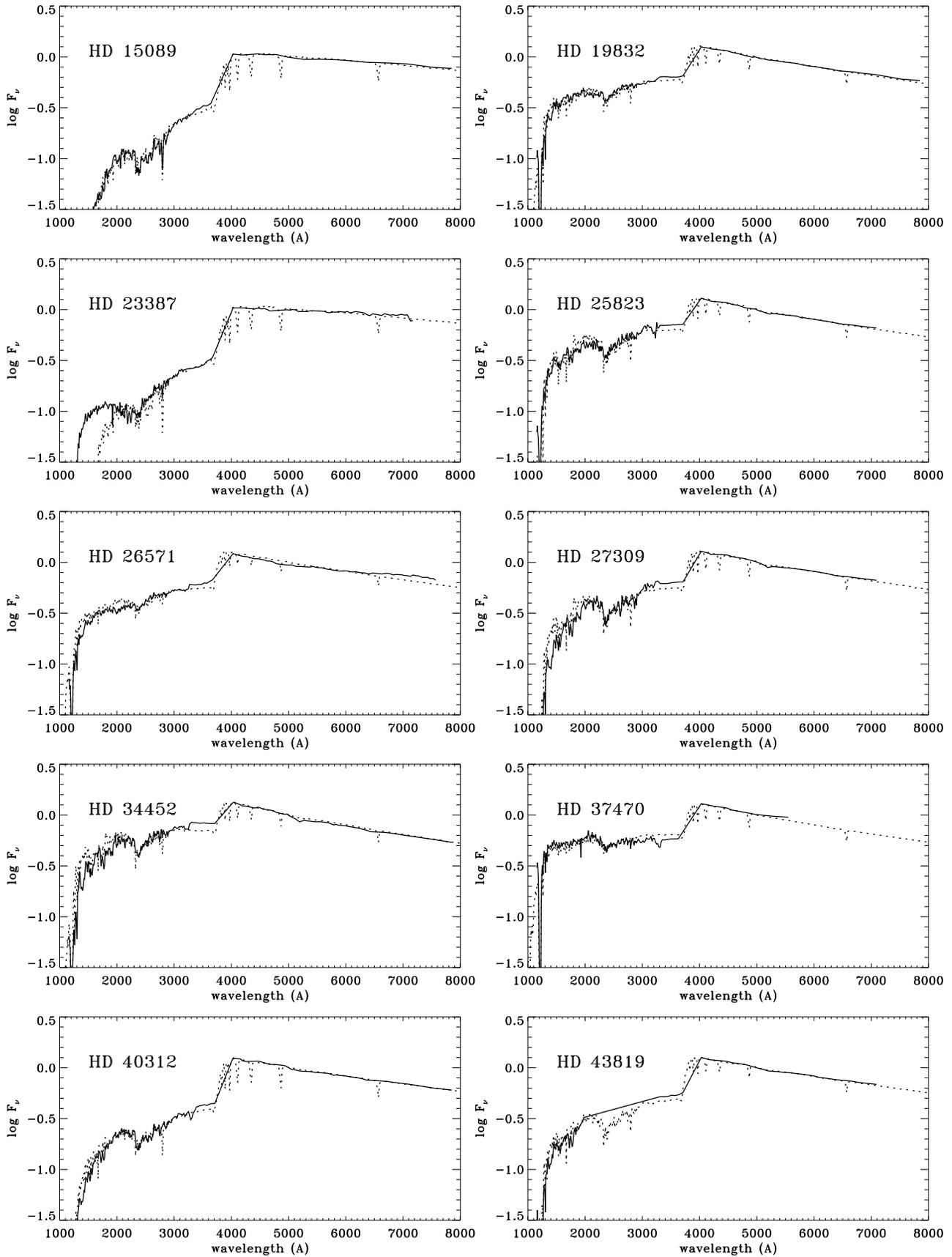}}}
\caption{\label{fits}
The observed SED (solid lines) with the best fitting models (broken lines).}
\end{center}
\end{figure*} 

\setcounter{figure}{0}
\begin{figure*}
\begin{center}
\rotatebox{0}{\scalebox{0.85}{\includegraphics{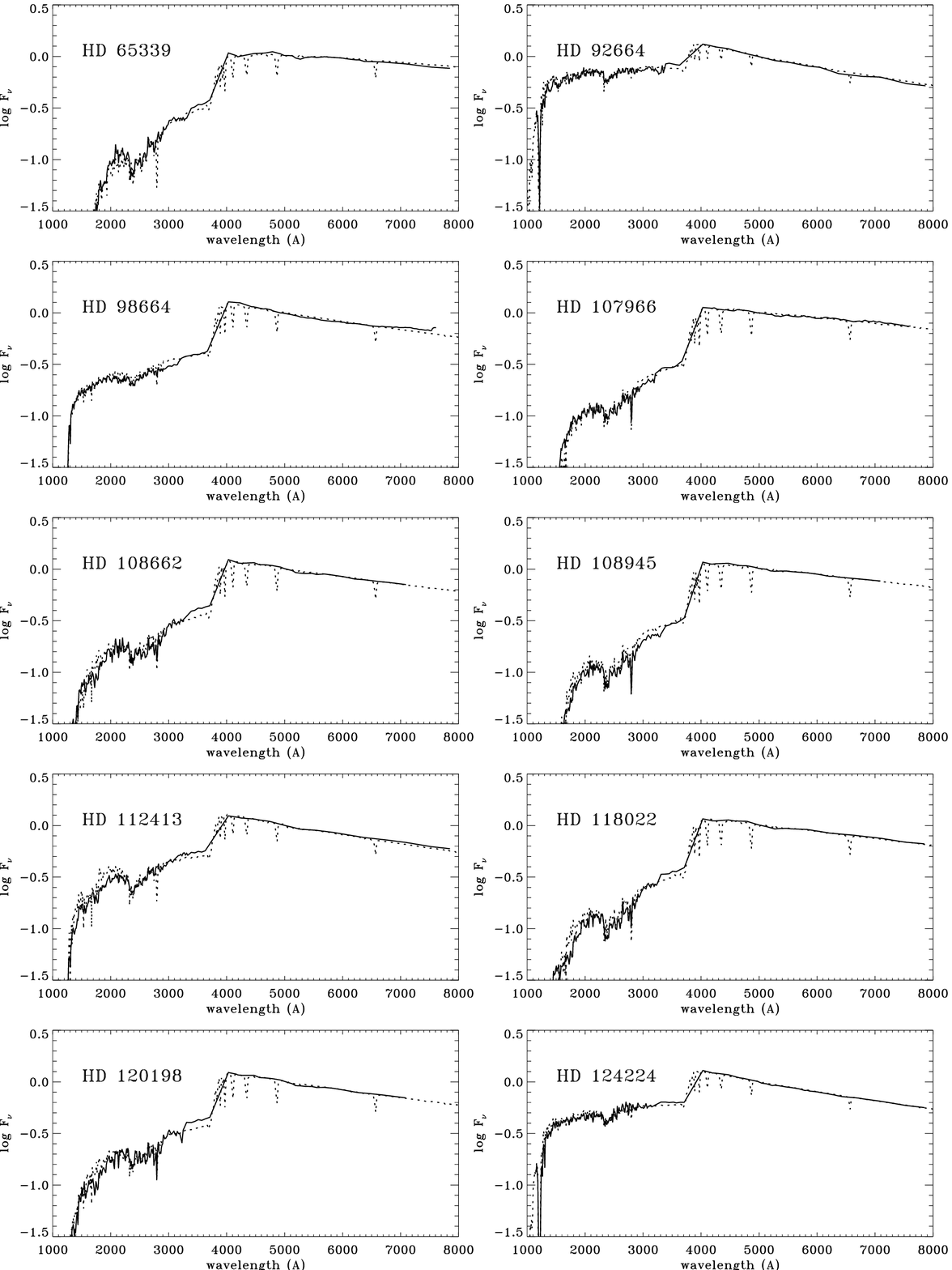}}}
\caption{ -- continued.}
\end{center}
\end{figure*} 

\setcounter{figure}{0}
\begin{figure*}
\begin{center}
\rotatebox{0}{\scalebox{0.85}{\includegraphics{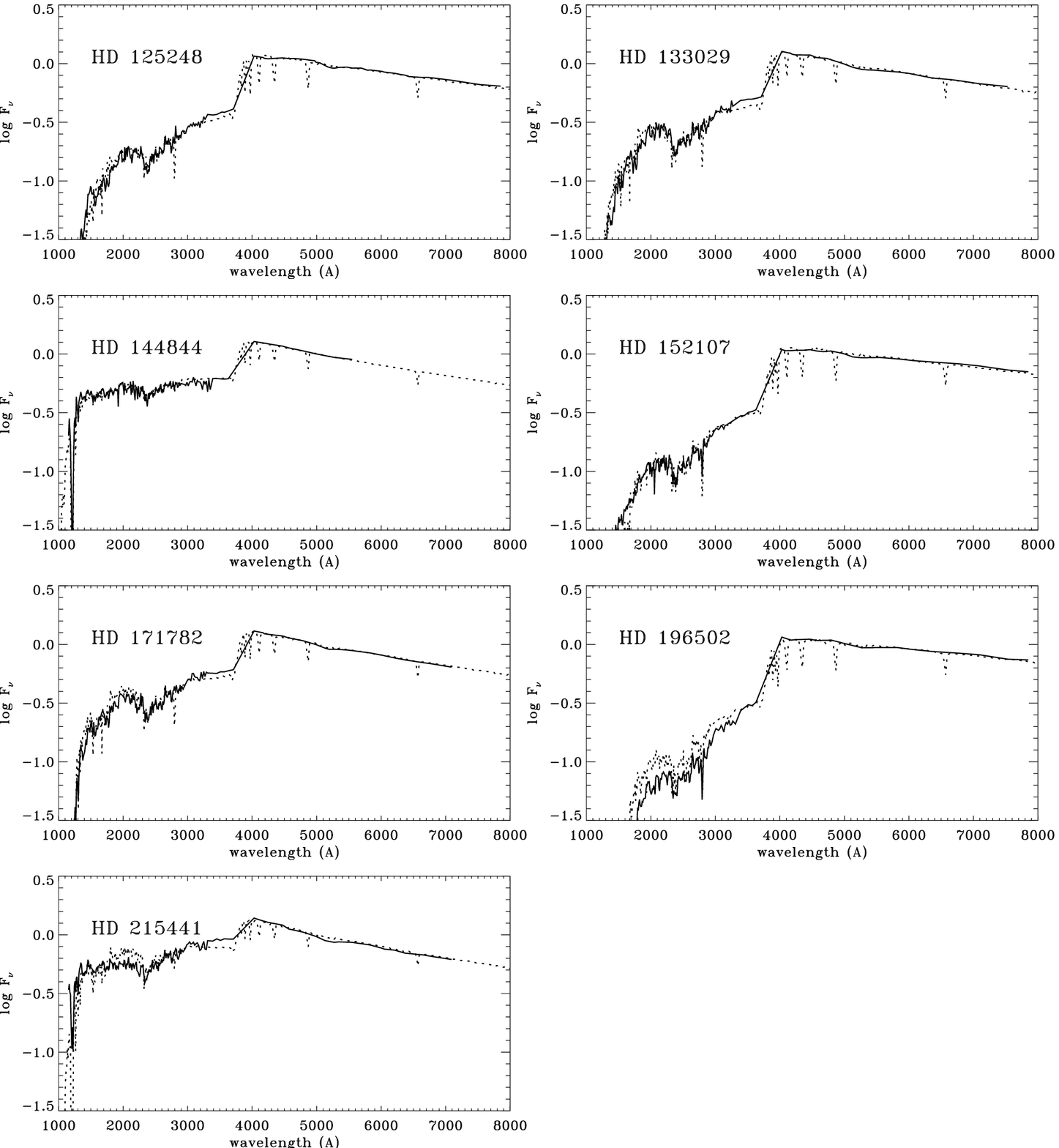}}}
\caption{ -- concluded.}
\end{center}
\end{figure*} 

\section{Results}

\subsection{Effective temperatures}

The root-mean-square (RMS) method was used to fit a theoretical SED to
observations of each star. Equal weights were given to the UV and visual
part of SED. Because the numbers of matching points are different in both
spectral regions, weights of individual points were equal to
$n_{\rm{UV}}^{-1}$ and $n_{\rm{vis}}^{-1}$, where $n_{\rm{UV}}$ and
$n_{\rm{vis}}$ are, respectively, the numbers of matching points in the UV and
visual scan of the considered star.

If \{$f_i$\} denote the observed values of the flux and \{$F_i$\} the
theoretical flux values from the model, we look for a (linear) factor $a$
such that ${\bf f} = a{\bf F}$ gives the best RMS fit. The root mean square
error (RMSE) $\sigma$ is a measure of the quality of fit

\begin{equation}
\sigma = \sqrt{\frac{\sum(f_i - aF_i)^2}{n - 1}}\,,
\end{equation}

where $n$ is a number of fitting points. A theoretical SED depends on three
basic parameters: effective temperature, metallicity and surface
gravity. It is known, however, that the effective temperature is the
primary parameter influencing SED. Variations of SED due to variations of
gravity or metallicity are of the second order (compared to temperature
variations). Moreover, they are correlated in a sense that a change of the
model SED due to decreased metallicity can be compensated by the increased
gravity (at constant effective temperature), and {\it vice versa}. If
reddening is treated as an additional free parameter, our calculations show
that there exists a trade-off between it and the stellar parameters. As a
result, the absolute minimum of RMSE may be spurious and a secondary
minimum should be considered as more reliable. That happened to HD 215441
(see below).

Even with reddening fixed, a simultaneous determination of temperature,
metallicity and gravity is not possible, at least in case of mCP stars. Surface
gravity of these stars is usually determined from other observations,
e. g. profiles of Balmer lines or $\beta$-index \citep{AR00}.
Unfortunately, the Balmer line profiles in mCP stars are subject to
modifications due to the Lorentz force influencing the effective gravity in
high atmospheric layers \citep{Ste78}. As a result, their intensity often
varies with rotation phase \citep{MM88}. Gravity determined from such lines
is more uncertain than in case of non-magnetic stars. All available
observational data indicate, nevertheless, that most, if not all, mCP stars
are main sequence (MS) objects, which means that their values of $\log g$
concentrate around 4.0 with a scatter of about $\pm$ 0.3-0.4 (see
e. g. \citealt{AR00}). We assumed therefore $\log g$ = 4.0 for all
investigated stars except two giants, HD 26571 and HD 43819, for which
$\log g$ = 3.0 was adopted based on their spectral class. It should be
stressed, nevertheless, that the models with $\log g$ = 4.0 and metallicity
increased by 0.5dex fit the observations of both giants as well as the
accepted models. Fortunately, the obtained effective temperatures are
insensitive to such variations of model parameters.

With gravity fixed, the effective temperature and metallicity were varied
when fitting models to observations, i.\,e. ${\bf F} = {\bf
F}(T_e,{\rm{m/H}})$. The best fitting values of the parameters were
determined from the $\sigma$ minimum. A typical minimum was rather broad -- the
change of metallicity by 0.5 dex and/or effective temperature by 250 K
resulted in a change of RMSE by 5-10 \%, which shows that the actual
accuracy of the optimum values of $T_e$ and [m/H] is of that order. Eleven
stars are common in Ad+Br and ABGK catalogs (see Table~\ref{list}). For
them separate fits were obtained to visual scans taken from the respective
catalogs (without IUE scans). Table~\ref{common} lists the results. The
average difference between temperatures based on both groups of catalogs is
about $-$70 K which is negligible. We conclude that no systematic
difference exists between spectrophotometric data from Ad+Br and ABGK
catalogs. The latter catalogs were used together with IUE data to determine
the effective temperatures of HD 23387, 26571, 98664 and 107966.

HD 215441 was treated in a special way. The best fit to its full SED
corrected for the interstellar reddening $E(B-V)$ = 0.20 \citep{Ste68}
resulted in metallicity of 1.0 and the effective temperature of 13 000
K. However, the quality of fit was extremely poor. The theoretical model
showed a substantially higher Balmer jump than observations and a different
slope of the Paschen continuum. Also the UV part of SED was poorly
reproduced. In addition, a fit of theoretical models separately to the
visual part of SED and then to the UV part of SED gave the temperatures
equal to 14 750 K and 14 000 K, respectively, i.e. significantly higher
than that resulting from the best fit to the full SED (see below). We
decided therefore to treat the reddening as an additional free parameter
and to determine it from the best fit together with metallicity and
effective temperature.  The procedure turned out to be only partly
satisfactory. The lowest value of RMSE (normalized to the total flux) was
obtained for [m/H] = 0.5, $T_e$ = 16 000 K and $E(B-V)$ = 0.37. We think
that such values are unreliable and they result most likely from a
correlation between the influence of [m/H] and $E(B-V)$ on SED (similarly
as it is in case of [m/H] and $\log$ g). Because of that, both parameters
cannot be simultaneously determined together with the effective
temperature from the SED alone. To get rid of degeneracy we {\em adopted}
[m/H] = 1.0 and looked then for minimum of RMSE by varying effective
temperature and reddening only. The best fit was obtained for $E(B-V)$ =
0.26 and $T_e$ = 14 000 K. We consider the fit as physically feasible and
we accepted it as our final result although some discrepancies still remain
(see Fig.~\ref{fits}).

Table~\ref{final} lists the final temperatures and metallicities resulting
from fits to full SEDs and Fig.~\ref{fits} shows the observed SEDs with the
best fitting models. With known $T_e$, angular diameters of the investigated
stars were calculated from the equation

\begin{equation}
\theta = \sqrt{\frac{4f_t}{\sigma T_e^4}}\,,
\end{equation}

where $\sigma$ is the Stefan-Bolzmann constant and $f_t$ is the integrated
observed flux. From angular diameters and distances stellar radii
in solar units can be obtained (Table~\ref{final}).

\begin{figure}
\begin{center}
\rotatebox{0}{\scalebox{0.42}{\includegraphics{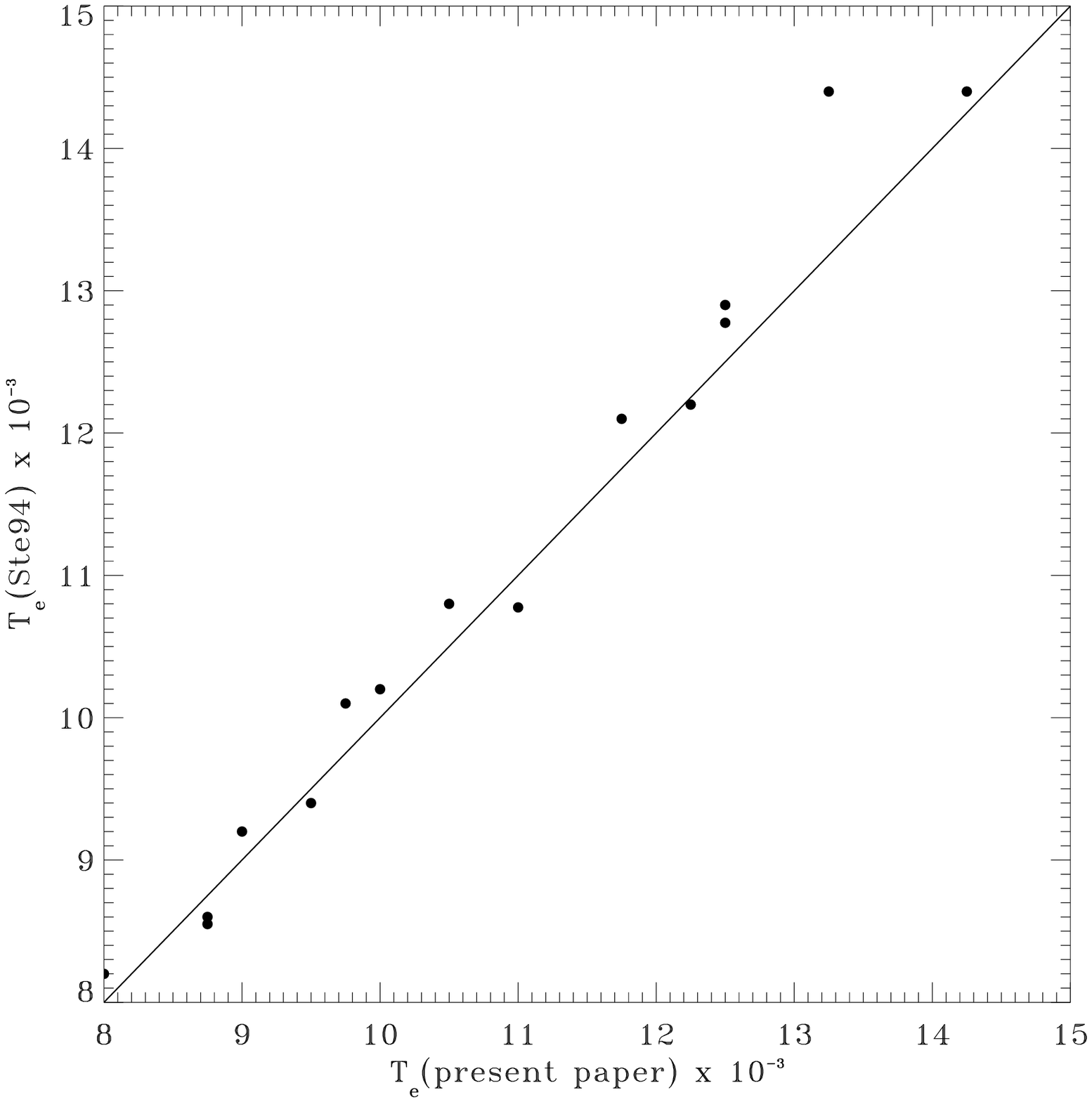}}}
\caption{ \label{compar}
A comparison of effective temperatures from the present paper with
temperatures listed by \citet{Ste94}.}
\end{center}
\end{figure}

Earlier determinations of the effective temperatures of mCP stars, based on
the observed SED were collected and discussed by \citet{Ste94}. 15 stars
from that paper are in common with the present investigation.
Fig.~\ref{compar} compares effective temperatures of these stars. Diagonal
is shown as a solid line. As can be seen from the figure, older
determinations agree satisfactorily with the present ones but a small
systematic difference between them appears in a sense that the older
temperatures are on average hotter by 190 K. The difference is within the
uncertainty of the temperature determination and probably reflects the
slightly better description of the metal opacity in UV by the recent
models, compared to earlier ones.

\subsection{Uncertainties}

\begin{figure}
\begin{center}
\rotatebox{0}{\scalebox{0.38}{\includegraphics{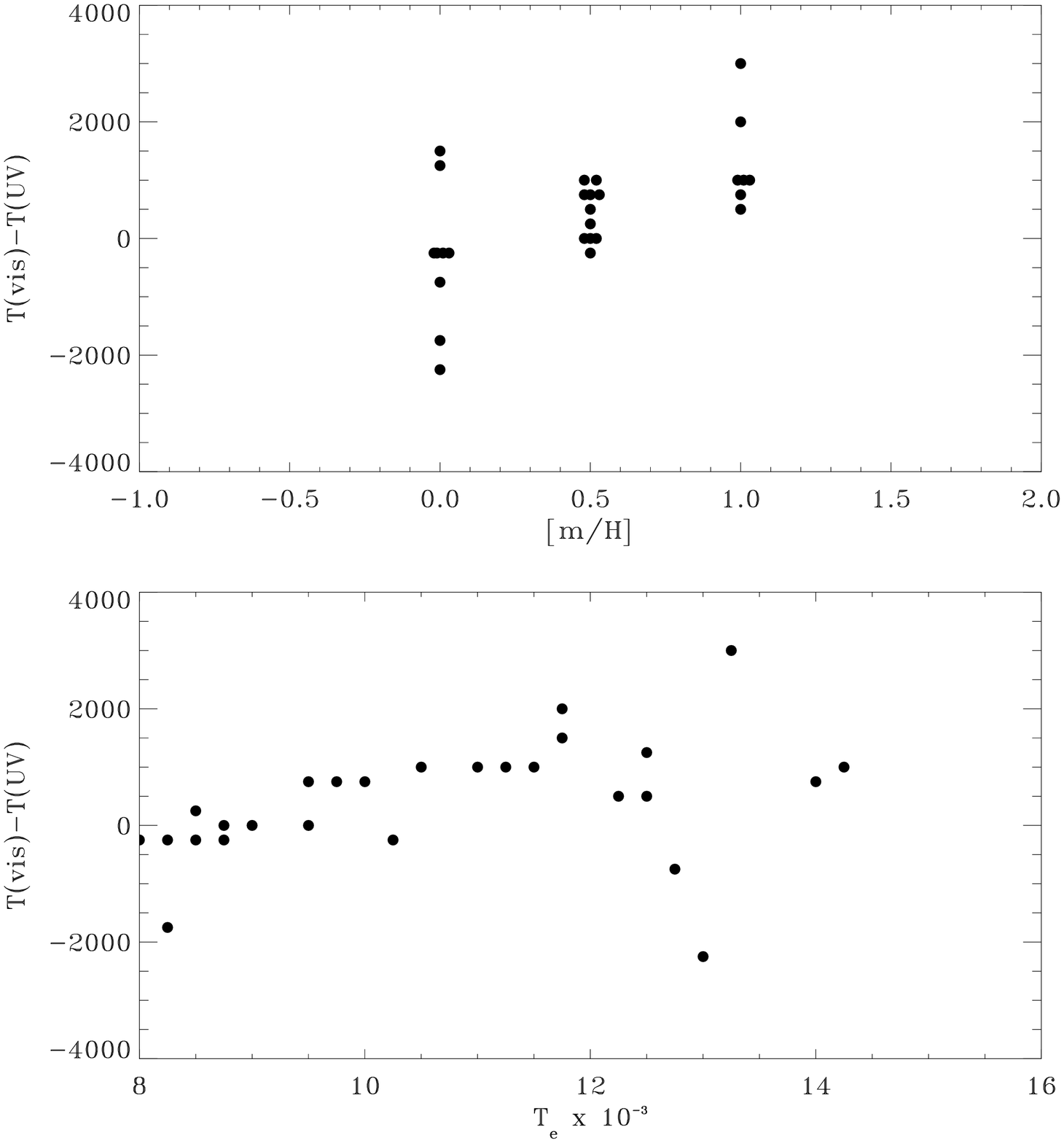}}}
\caption{\label{deltate}
The dependence of $\Delta T = T$(vis) - $T$(UV) on metallicity (top) and
effective temperature (bottom).}
\end{center}
\end{figure}

As it was mentioned earlier, the uncertainty of the best fit, estimated
from the shape of the minimum of RMSE is of the order of 200-300 K.  It can
be seen from Fig.~\ref{fits}, however, that while in some cases the
observed SEDs agree very well with the best fitting models, there are also
cases where the fit is rather poor (e. g. HD 34452 or 215441). Uncertainty
of each individual determination of the effective temperature should
reflect this quality of fit. To measure it quantitatively, models were
fitted separately to the UV scan of each star and then to the visual
scan. Gravity and metallicity were the same as in the global best fitting
models to the full SEDs and were kept fixed. Table~\ref{final} shows the
results. In majority of cases temperature found from the visual scan is
higher than from the UV scan but in several stars it is lower. The
difference $T({\rm{vis}}) - T({\rm{UV}}) = \Delta T$ is plotted versus
$T_e$ and [m/H] in Fig.~\ref{deltate}. Note that any possible uncertainties
of UV calibration relative to the visual observations (suggested by
apparent gaps between both scans visible in some stars) have no influence
on thus found temperatures - only the shape of SED matters. In a case of
the perfect fit between the theoretical model and observations, separate
fits to UV and visual scan should give identical temperatures or, at most,
differing by a value of the estimated error from the best fit to the full
SED which, in nearly all cases, lies between $T({\rm{vis}})$ and
$T({\rm{UV}})$. That means that the difference $|\Delta T| \le$ 500 K
(i.\,e. twice the error) indicates a satisfactory quality of fit. This
occurs in 11 stars. In addition, there are 5 ``border cases'' with $|\Delta
T|$ = 750 K. In all other stars it exceeds that value and the most extreme
differences reach -2250 K and 3000 K.  Such large differences indicate that
the best fitting models do not yet give satisfactory results. A closer
inspection of SEDs of stars with apparently close effective temperatures,
e. g. HD 34452 and HD 37470 (Fig.~\ref{fits}), reveals, indeed, profound
differences in shapes of UV spectrum, which result from a different UV line
blanketing between both stars. Obviously, scaled solar abundances are a
very poor alternative of the true chemical compositions.  The largest
positive values of $\Delta T$ are present in stars with the highest
available metallicity as can be seen from Table~\ref{final} and
Fig.~\ref{deltate}. This indicates that the solar metallicity enhanced by a
factor of 10 is still insufficient to correctly reproduce the observed
SED. Perhaps models with [m/H] = 1.5 or even 2.0 would do it better. Note
that \citet{Lec74} adopted line opacity enhanced by a factor of 100 to
qualitatively reproduce differences between normal and Ap star flux
distributions. Substantial discrepancies between $T$(UV) and $T$(vis) do
not necessarily mean that the present effective temperatures are in gross
error. Future models with individually adjusted chemical compositions will
simply narrow an existing gap between these two temperatures, although they
may close the gap at a somewhat different value of effective temperature
than the present one. Large negative differences appear for two stars: HD
23387 and HD 37470. The former star shows unusually flat Paschen continuum
which results in low $T$(vis). Such flat or, sometimes, rounded continua with a
depression at the short wavelength end appear in a few  Ap
stars during all, or some of the phases of the rotation period
\citep{Kod73, PA83} due to a very strong blanketing effect concentrated in
this spectral region. That makes the Paschen continuum of such stars look
unusually flat. This does not seem, however, to be the case for HD 23387
which is, at most, mildly peculiar, or may even be normal \citep{Lan07}. At
present we cannot offer a convincing explanation for the existing
discrepancy. Obviously, new, more accurate observations of this star are
needed. The other star, HD 37470, was observed in visual up to 5500 A
only. In addition, the red end of the Paschen continuum looks unusual (see
Fig.~\ref{fits}), so it is possible that the uncertain visual scan is
responsible for the discrepancy.

In one star, HD 25823, the best fitting temperature is {\em lower} than
either $T(\rm{UV})$ or $T(\rm{vis})$. This unexpected result is most likely
due to a particularly large mismatch of both parts of SED. If the UV SED is
too faint compared to visual SED, the best fitting model will also give a
too low temperature compared to the temperatures resulting from separate
fits to UV and visual parts of SED. Conversely, too bright UV SED compared
to the visual SED results in a too high best fitting value of the
temperature. For HD 25823 the best fitting temperature is lower by 250 K
than $T(\rm{UV})$ and by 750 K than $T(\rm{vis})$, so that all three
temperatures are within 2-3 $\sigma$ from each other. In several other
stars the best fitting temperature is very close to, or identical with one
of the temperatures found from the UV or visual scan. This may also be the
result of a small mismatch of both parts of SED.

An additional source of uncertainty of effective temperatures of the
investigated stars is connected with poorly known reddening
corrections. The accepted $E(B-V)$ values may in some cases be in error up
to a few hundredths of magnitude. The best fitting temperature increases by
about 1 \% per 0.01 mag increase of $E(B-V)$ as our numerical experiment
showed. Assuming that $E(B-V)$ may be in error up to 0.03 mag we obtain the
value of about 250 - 400 K for the maximum error of effective temperature
resulting from poorly known reddening. 

{\bf Variability, ubiquitously observed in mCP stars, may result in
  apparent temperature variations with rotational phase \citep{MS80,
  SM87}. Our final temperatures can be affected by such variations,
  particularly for stars with few scans unfavorably distributed in
  phase.} 

As it is seen, all the discussed error sources have a highly
  various, and difficult to assess, importance for individual
  stars. Therefore, instead of trying to determine the uncertainty for each
  star separately, we estimate an {\em average} uncertainty at about 500 K.

\begin{table*}
\caption[]{Final results.}
\label{final}
\begin{tabular}{rcrcrrrrccc} 
\hline
HD & [m/H] & $T_e$(final) & $\Theta$ & $BC$ & $T$(vis) & $T$(UV)& $\Delta T$ 
& $f_t\times 10^7$ &$\theta$(mas) & $R/R_{\odot}$  \\
\hline
15089 & 0.0 & 8250 & 0.611 &-0.01 &8250 & 8500 & -250 & 3.86 &0.50 & 2.3 \\
19832 & 0.5 & 12250 & 0.411 &-0.74 &12750 & 12250 & 500 & 2.43 &0.18 & 2.1 \\
23387 & 0.0 & 8250 & 0.611 &-0.05 &8000 & 9750 & -1750 & 0.347 &0.15 & 2.1 \\
25823 & 1.0 & 12500 & 0.403 &-0.67 &13250 & 12750 & 500 & 3.94 &0.22 & 3.6 \\
26571 & 0.0 & 11750 & 0.429 &-0.76 &12250 & 10750 & 1500 & 3.66 &0.24 & 8.1 \\
27309 & 1.0 & 11750 & 0.429 &-0.58 &12500 & 10500 & 2000 & 3.07 &0.22 & 2.3 \\
34452 & 1.0 & 13250 & 0.380 &-0.80 &14750 & 11750 & 3000 & 3.71 &0.19 & 2.9 \\
37470 & 0.0 & 13000 & 0.388 &-0.96 &12250 & 14500 & -2250 & 0.466 &0.07 & 1.9 \\
40312 & 0.5 & 10000 & 0.504 &-0.26 &10500 & 9750 & 750 & 28.2 &0.92 & 5.2 \\
43819 & 0.5 & 11000 & 0.458 &-0.38 &11250 & 10250 & 1000 & 1.10 &0.15 & 3.2 \\
65339 & 0.0 & 8000 & 0.630 &-0.02 &8000 & 8250 & -250 & 0.995 &0.27 & 2.8 \\
92664 & 0.5 & 14250 & 0.354 &-1.01 &15250 & 14250 & 1000 & 3.97 &0.17 & 2.7 \\
98664 & 0.0 & 10250 & 0.492 &-0.28 &10250 & 10500 & -250 & 7.78 &0.46 & 3.3 \\
107966 & 0.0 & 8500 & 0.593 &-0.07 &8500 & 8750 & -250 & 2.25 &0.36 & 3.3 \\
108662 & 0.5 & 9500 & 0.531 &-0.15 &10250 & 9500 & 750 & 2.28 &0.29 & 2.6 \\
108945 & 0.5 & 8750 & 0.576 &0.01 &8750 & 8750 & 0 & 1.64 &0.29 & 3.0 \\
112413 & 1.0 & 11250 & 0.448 &-0.42 &11750 & 10750 & 1000 & 25.4 &0.69 & 2.5 \\
118022 & 0.5 & 9000 & 0.560 &-0.05 &9000 & 9000 & 0 & 2.83 &0.36 & 2.2 \\
120198 & 0.5 & 9750 & 0.517 &-0.20 &10500 & 9750 & 750 & 1.59 &0.23 & 2.2 \\
124224 & 0.0 & 12500 & 0.403 &-0.79 &13000 & 11750 & 1250 & 5.08 &0.25 & 2.2 \\
125248 & 0.5 & 9500 & 0.531 &-0.17 &9500 & 9500 & 0 & 1.31 &0.22 & 2.1 \\
133029 & 1.0 & 10500 & 0.480 &-0.40 &11250 & 10250 & 1000 & 1.04 &0.16 & 2.4 \\
144844 & 0.0 & 12750 & 0.395 &-0.77 &12750 & 13500 & -750 & 3.18 &0.19 & 2.7 \\
152107 & 0.5 & 8750 & 0.576 &-0.01 &8750 & 9000 & -250 & 2.97 &0.39 & 2.3 \\
171782 & 1.0 & 11500 & 0.438 &-0.53 &12000 & 11000 & 1000 & 0.472 &0.09 & 2.7 \\
196502 & 0.5 & 8500 & 0.593 &-0.02 &8500 & 8250 & 250 & 2.13 &0.35 & 4.8 \\
215441 & 1.0 & 14000 & 0.360 &-0.80 &14750 & 14000 & 750 & 0.32 & 0.05 & 4.2 \\
\hline
\end{tabular}
\end{table*}

\subsection{Comments on individual stars.} 

Approximate values of variability periods $P_{\rm{rot}}$ and average
effective magnetic fields are taken from \citet{BBM03, BBM05}, unless
otherwise indicated.

{\it HD 15089}, $P_{\rm{rot}}$ = 1.74 d and $\overline{B_e}$ = 200 G.

{\it HD 19832},  $P_{\rm{rot}}$ = 0.73 d and $\overline{B_e}$ = 300 G.

{\it HD 23387}, Pleiades member, unknown rotation period,
  $\overline{B_e}$ = 1.8 kG. \citet{Lan07} considers the star as
  normal with no magnetic field.

{\it HD 25823},  $P_{\rm{rot}}$ = 4.66 d and $\overline{B_e}$ = 700 G.

{\it HD 26571}, B9IIIp, reddening from \citet{Iso86},$P_{\rm{rot}}$ =
15.7 d, mag. field ?

{\it HD 27309},  $P_{\rm{rot}}$ = 1.11 d and $\overline{B_e}$ = 1.8 kG.

{\it HD 34452},  $P_{\rm{rot}}$ = 2.47 d and $\overline{B_e}$ = 700 G.

{\it HD 37470}, reddening from \citet{SC71}, unknown rotation
period, $\overline{B_e}$ = 160: G.

{\it HD 40312}, $P_{\rm{rot}}$ = 3.62 d and $\overline{B_e}$ = $\sim$ 500
G, radius $R = 3.6 R_{\odot}$ \citep{BL80}.

{\it HD 43819}, B9IIIp, $P_{\rm{rot}}$ = 15.03 d \citep{AY05},
$\overline{B_e}$ = 300 G.

{\it HD 65339}, $P_{\rm{rot}}$ = 8.03 d and $\overline{B_e}$ = 3 kG, binary
with $P_{\rm{bin}}$ = 2400 d. Nearly identical components with masses
$M_{\rm{1,2}} = 1.5 M_{\odot}$ \citep{Car02}.

{\it HD 92664}, $P_{\rm{rot}}$ = 1.67 d and $\overline{B_e}$ = 800 G. 

{\it HD 98664}, unknown rotation period, mag.field?, normal star?

{\it HD 107966}, member of Coma cluster, SB?, $P_{\rm{rot}} >$ 1 d
\citep{Jac72}, mag. field?, Am?

{\it HD 108662}, member of Coma cluster, blue straggler?, $P_{\rm{rot}}$ =
5.08 d and $\overline{B_e}$ = 600 G, mass $M = 2.4 M_{\odot}$ \citep{Cas06}.

{\it HD 108945}, member of Coma cluster, $P_{\rm{rot}}$ = 1.92 d and
$\overline{B_e}$ = 500 G, mass $M = 2.4 M_{\odot}$ \citep{Cas06}.

{\it HD 112413},  $P_{\rm{rot}}$ = 5.47 d and $\overline{B_e}$ = 1.3 kG.

{\it HD 118022},  $P_{\rm{rot}}$ = 3.72 d and $\overline{B_e}$ = 800 G.

{\it HD 120198}, $P_{\rm{rot}}$ = 1.39 d and $\overline{B_e}$ = 700 G.

{\it HD 124224},  $P_{\rm{rot}}$ = 0.52 d and $\overline{B_e}$ = 600 G.

{\it HD 125248},  $P_{\rm{rot}}$ = 9.30 d and $\overline{B_e}$ = 1.5 G, $R
= 1.97 R_{\odot}$ \citep{Bag02}.

{\it HD 133029},  $P_{\rm{rot}}$ = 2.11 d and $\overline{B_e}$ = 2.5 G.

{\it HD 144844},  reddening from \citet{Gor94}, unknown rotation
period, $\overline{B_e}$ = 300 G.

{\it HD 152107}, UMa moving group, SB with a faint secondary
\citep{King03}, $P_{\rm{rot}}$ = 3.87 d and $\overline{B_e}$ = 500 G.

{\it HD 171782},  reddening from \citet{Ade80}, $P_{\rm{rot}}$ = 4.47 d
\citep{PM98}, mag. field?

{\it HD 196502},  $P_{\rm{rot}}$ = 20.3 d and $\overline{B_e}$ = 500 G.

{\it HD 215441}, reddening from the present paper, mass $M = 3-4 M_{\odot}$
\citep{Lan89}, $P_{\rm{rot}}$ = 9.49 d, $\overline{B_e}$ = 17 kG.

\begin{figure}
\begin{center}
\rotatebox{0}{\scalebox{0.38}{\includegraphics{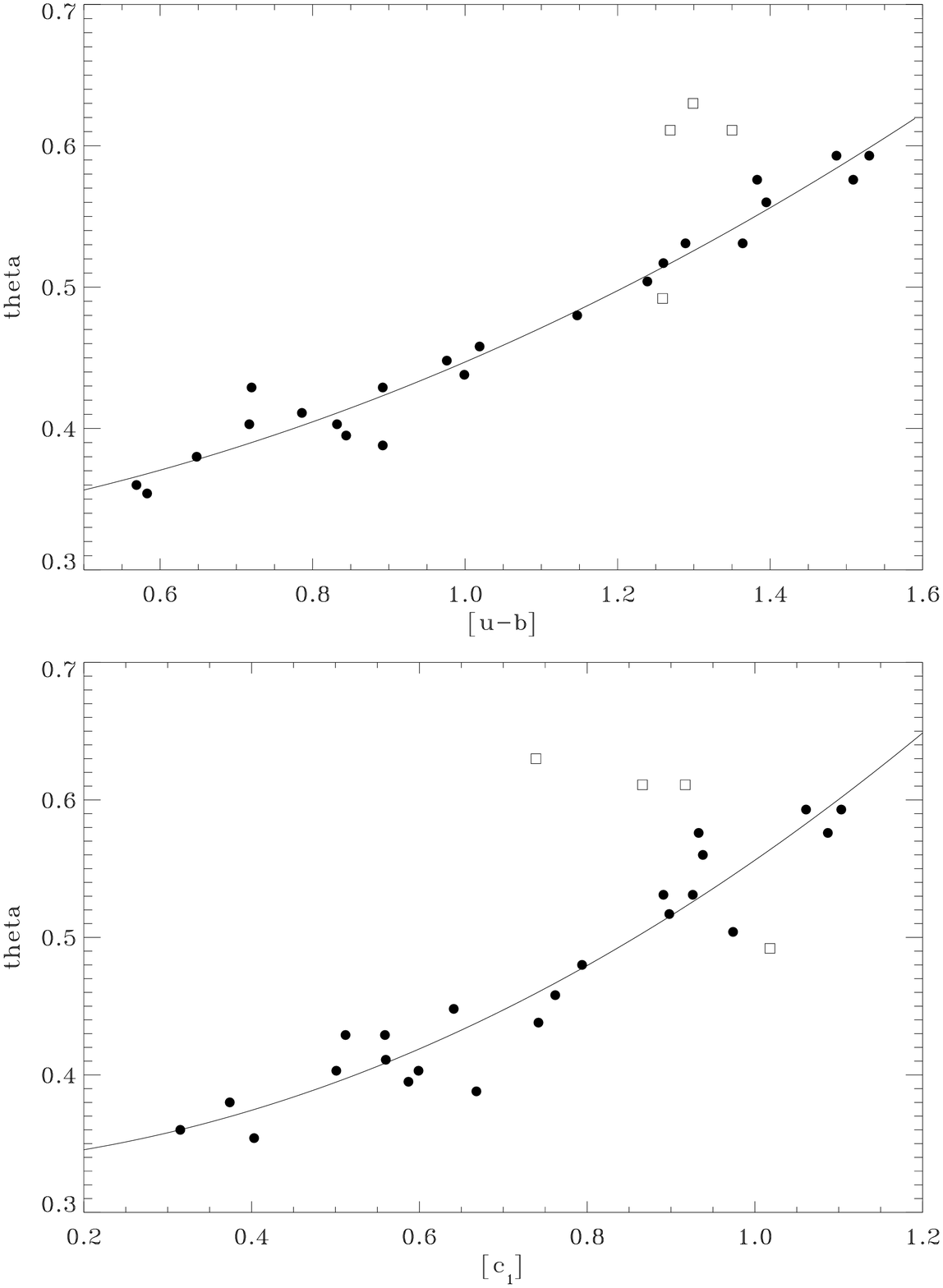}}}
\caption{\label{stromcal}
A new calibration of Str\" omgren reddening free photometric indices in
terms of effective temperature of the MCP stars. Best fitting lines are
described in the text.}
\end{center}
\end{figure}

\begin{figure}
\begin{center}
\rotatebox{0}{\scalebox{0.38}{\includegraphics{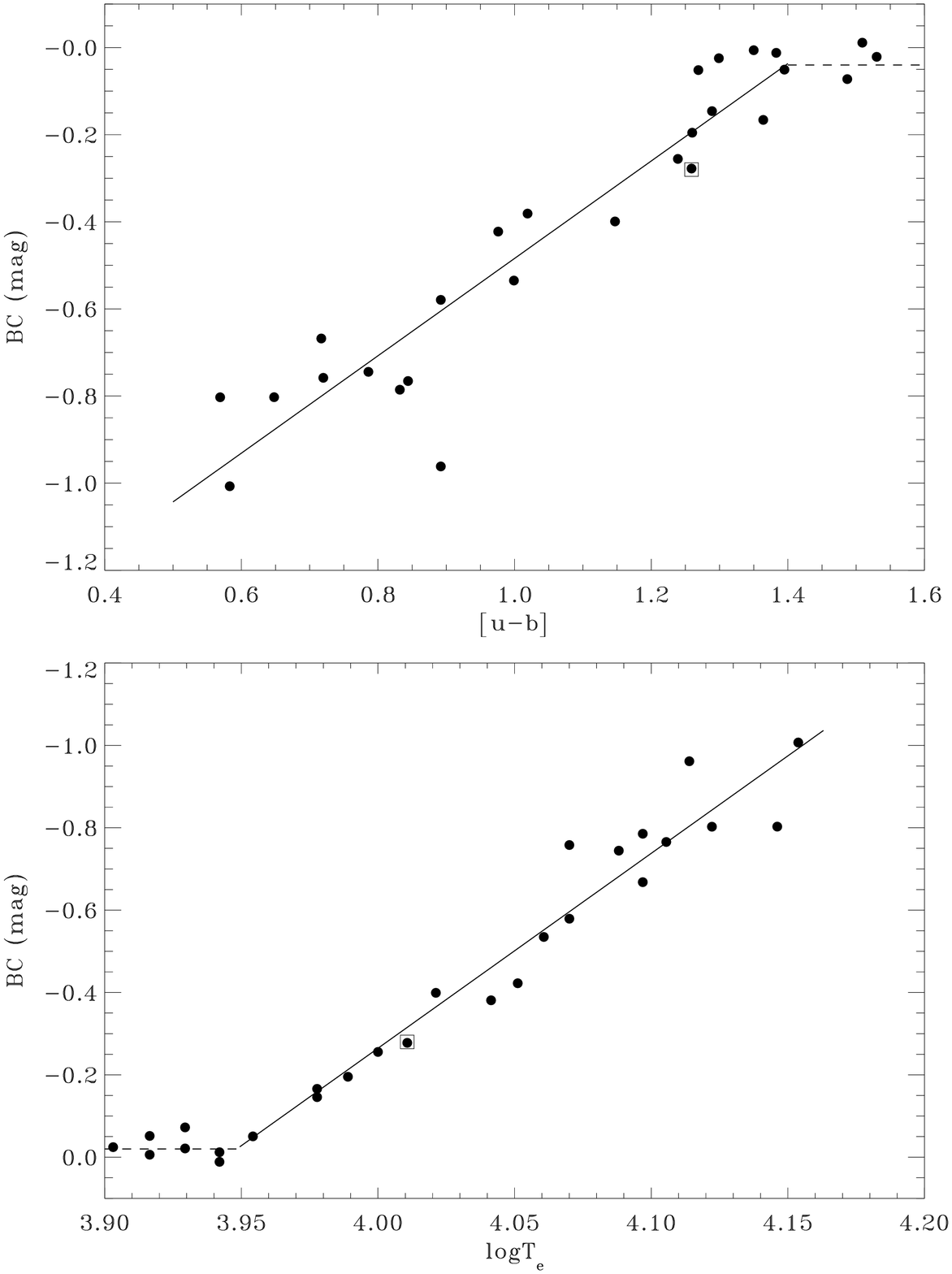}}}
\caption{\label{bol}
A new calibration of bolometric correction for MCP stars in terms of
reddening free index $[u-b]$ (top) and effective temperature (bottom). Best
fitting lines are described in the text.}
\end{center}
\end{figure}

\subsection{Revised photometric calibrations}

All the analyzed stars have been measured in the Str\" omgren photometric
system so reddening free indices $[u-b]$ and $[c_1]$ can be
calculated for them (see Table~\ref{list}). They are defined as

%\begin{eqnarray}
%[u-b] & = & (u-b) - 1.53(b-y),\\
%[c] & = & c_1 - 0.20(b-y). 
%\end{eqnarray}

\begin{equation}
[u-b]  = (u-b) - 1.53(b-y)\,,
\end{equation}

\begin{equation}
[c_1]  =  c_1 - 0.20(b-y)\,.
\end{equation}

Figure~\ref{stromcal} presents the dependence of the parameter $\Theta =
5040/T_e$ on $[u-b]$ (top) and on $[c_1]$ (bottom). The data clearly show
that quadratic fits are required to describe satisfactorily both
dependencies. The calculations confirm indeed that the quadratic terms are
significant. Four stars, marked with squares, were excluded from the best
fits. Three high lying symbols correspond to three coolest stars from the
sample, with temperatures below 8500 K. Both Str\" omgren indices are
sensitive to the height of the Balmer jump, hence they are good measures of
temperature only for stars hotter than about 8500 K, i.e. as long as the
height of the Balmer jump increases monotonically with decreasing temperature. Beyond
this temperature the jump {\em decreases} with a decreasing temperature,
which is reflected by an apparent shifting of such stars on the
index-temperature diagrams towards lower values of the indices as visible
in the Fig.~\ref{stromcal}. Fourth star, HD 98664, is apparently a
normal star (see the comments above).

The best fit to the data in the $[u-b]$ - $\Theta$ diagram is given by

\begin{eqnarray}
\Theta &=& 0.3167 + 0.0284[u-b] + 0.1017[u-b]^2\,, \\
&&0.55<[u-b]<1.55\,. \nonumber
\end{eqnarray}

Earlier calibrations were linear in $[u-b]$ so, for the sake of comparison,
we also fitted a straight line to the data. The result is

\begin{equation}
\Theta = (0.2121\pm 0.0131) + (0.2442\pm 0.0120)[u-b]\,,
\end{equation}

We give four significant digits for the straight comparison of our result 
with \citet{Nap93}, although it is clearly seen that the last digits are
meaningless. Their relation 

\begin{equation}
\Theta = 0.2162 + 0.2301[u-b]\,,
\end{equation}

and the relation derived by \citet{Ste94}

\begin{equation}
\Theta = 0.200 + 0.246[u-b]\,,
\end{equation}

differ insignificantly from the presently derived linear fit. It means that
both older calibrations are not  grossly erroneous. The quadratic fit
refines, however, the existing calibrations.

Similarly, the best quadratic fit to the data in the $[c_1]$ - $\Theta$
diagram is given by

\begin{eqnarray}
\Theta &=& 0.3328 + 0.0237[c_1] + 0.1996[c_1]^2\,, \\
&&0.3<c_1<1.1\,. \nonumber
\end{eqnarray}

The linear fit

\begin{equation}
\Theta = (0.2382\pm 0.0179) + (0.3136\pm 0.0233)[c_1]\,,
\end{equation}

can be compared with \citet{Nap93}

\begin{equation}
\Theta = 0.2489 + 0.2698[c_1]\,.
\end{equation}

From $f_t$ one can find the apparent bolometric magnitude
$m_{\rm{bol}}$ of a star using the relation given by \citet{Code76} 

\begin{equation}
m_{\rm{bol}} = -2.5\log f_t - 11.51\,.
\end{equation}

The difference between $m_{\rm{bol}}$ and the apparent stellar visual
magnitude (corrected for reddening) gives the bolometric correction
$BC$. The bolometric corrections for the investigated stars are listed in
Table~\ref{final} and plotted in Fig.~\ref{bol} as a function of $[u-b]$
(top) and effective temperature (bottom). The best fitting relations are
also plotted. Because HD 98664 lies very close to the best fitting lines in
both diagrams, it has a negligible influence on the best fits. It is marked
with a square overplotted on a solid circle and was included in the
fit. Solid lines give best linear fits to the data excluding the coolest
stars which all have $BC \approx 0$. A quadratic term in both cases turned
out to be insignificant. The solid lines are described by

\begin{eqnarray}
BC &=& (-1.602\pm 0.092) + (1.118\pm 0.086)[u-b]\, \\
&&0.5<[u-b]<1.4\,, \nonumber
\end{eqnarray}

and by

\begin{eqnarray}
BC &=& (18.668\pm 0.975) - (4.733\pm 0.241)\log T_e\,, \\
&&3.94<\log T_e <4.16\,. \nonumber
\end{eqnarray}

\section{Conclusions}

Effective temperatures were determined for 27 stars. All but HD 98664,
possess at least two properties characteristic of mCP stars, i.e. chemical
peculiarities, light variations and/or measurable surface magnetic
field. The temperatures were found from a fit of a metal enhanced model
atmosphere to the full SED, including UV spectrophotometry from IUE and
visual scan from ground based observations. In case of distant stars the
observed SEDs were corrected for reddening before searching the best
fitting model. Our experience with fitting models to observations showed
that reliable results can be obtained only when at most two free parameters
describing a model are used in the best fit, e.g. metallicity and
temperature. Introducing the third unknown parameter, e.g. reddening or
gravity, to be determined from the best fit simultaneously with the other
two produces either indeterminacy or gives a spurious result. We assumed
therefore $\log g$ = 4 for all main sequence stars and $\log g$ = 3 for two
known giants. We also adopted individual values of $E(B-V)$ from literature
or assumed zero reddening for close stars. In one case of HD 215441 we
attempted to determine the value of $E(B-V)$ from the best fit. However, to
obtain meaningful results we were forced to keep metallicity fixed when
fitting models to the observations.  In addition to fitting a model to the
full SED we also fitted a model separately to the UV and the visual part of
SED of each star. A difference between thus found temperatures is a measure
of quality of fit to the full SED. In about a half of stars the difference
is small enough to consider the fit acceptable but in the remaining stars
the difference was $\ge$ 1000 K which indicates a necessity to
improve the atmospheric models of mCP stars. The presently available
models reproduce the observations only approximately. Taking into account
all uncertainties, we estimate an error of determination of effective
temperature at 500 K.

Together with Str\" omgren photometry, new determinations of the effective
temperatures were used to produce improved color-temperature calibrations
for mCP stars. New data suggest quadratic calibration formulas in color
indices. From total integrated fluxes bolometric corrections were computed
and improved color-BC and temperature-BC calibrations were also determined.

\section{Acknowledgments}

KS acknowledges the partial financial support of the Ministry of Science
and Higher Education through the grant 1 P03 016 28.

\end{document}